# The CH$_3$CHOO 'Criegee intermediate' and its anion: Isomers, infrared spectra, and W3-F12 energetics

M. Kettner     A. Karton     A. J. McKinley     D. A. Wild[*]

*School of Chemistry and Biochemistry*
*The University of Western Australia*
*M310, 35 Stirling Hwy, Crawley 6009, Australia.*
DOI: 10.1016/j.cplett.2014.12.037 ARXIV: 1501.00353
*duncan.wild@uwa.edu.au

ABSTRACT: For the CH$_3$CHOO Criegee intermediates (ethanal-oxide) and analogous anions, we obtain heats of formations and electron affinities at CCSDT(Q)/CBS level of theory by means of the high-level W3-F12 thermochemical protocol. The electron affinities amount to 0.20 eV and 0.35 eV for the *cis* and *trans* isomer, respectively. Neutral *cis* and *trans* isomers are separated by 14.1 kJ mol$^{-1}$, the anions are almost isoenergetic (0.4 kJ mol$^{-1}$ separation). Harmonic vibrational frequencies are presented at CCSD(T)/aug'-cc-pVTZ level of theory. Since the synthesis of these species in gas-phase experiments might be possible in the near future, we include a predicted photoelectron spectrum.

## Introduction

Carbonyl oxides, also known as Criegee Intermediates (CI), are understood to be linking structures in tropospheric ozonolysis reactions.[1] During collisions between unsaturated hydrocarbons and ozone in the earth's atmosphere, these short-lived, zwitterionic molecules provide a pathway to the formation of ·OH radicals through unimolecular decomposition.[2–5] As one of the main gas-phase oxidants in our atmosphere, it is also likely to react with other carbonyl oxides as well as H$_2$O, SO$_2$, NO, NO$_2$, other alkenes or ozone molecules,[6–10] only to name a few examples for the rich chemistry these molecules exhibit.

It is this multitude of reactions that raises our curiosity towards such a rather unstable molecule, recently fuelled even more so by the discovery of alternate methods for gas-phase synthesis for the simplest of all CIs, the methanal-oxide (MO).[7,11] In one of these reactions, Welz and co-workers showed that CH$_2$OO can be obtained through photolysis of CH$_2$I$_2$ in the presence of O$_2$.[7] As a consequence, the behaviour of the intermediate could now be investigated utilising various spectroscopic methods, such as UV adsorption and gas-phase IR experiments;[12,13] yet due to its metastable nature, experiments on MO are indeed rather challenging. However, the small size of the intermediate allows for a thorough theoretical analysis.[14] For instance, Fang et al. studied the potential energy surface of the reaction of CH$_2$ and O$_2$ on the singlet and triplet surfaces using CASSCF-type calculations.[15] Nguyen et al.[16] obtained the heat of form-ation and ionisation energy of the Criegee intermediate at the CCSD(T)/CBS level using W1 theory.[17]

In natural atmospheric processes, the precursors to carbonyl oxide formation–volatile organic compounds (VOC)–are for example generated by our forests and as earth's biosphere slowly warms more VOCs can enter the carbonyl cycle.[18] The reaction of VOCs with ozone creates CIs with high internal energy, most of which immediately dissociate through unimolecular rearrangements leading to, amongst others, ·OH molecules.[14] CIs stable enough to participate in further reactions (lifetimes of a pprox. 100 ms were postulated), are often called stabilised Criegee Intermediates (sCI).[2,19]

The main focus of current and past research lies within the neutral CI's chemistry. However, it is also possible to explore the neutral CIs coming from an anionic surface via anion photoelectron spectroscopy. Last year, our group presented the first theoretical anionic structure for the simplest of all carbonyl oxides (MO) and found that it is stable with respect to dissociation.[20] Nakajima and Endo recently provided an experimental microwave spectrum of one of the ethanal-oxide isomers.[21] In light of these motivators, we discuss here the neutral and anionic properties of the ethanal-oxide. Due to its $\beta$-carbon, it seems more important to atmospheric processes than its 'simpler' counterpart.[5,18]

## Computational methods

The geometries and harmonic frequencies of the anion and neutral $CH_3CHOO$ species were obtained at the CCSD(T)/aug′-cc-pVTZ level of theory (where aug′ indicates the use of aug-cc-pVTZ on O and C and cc-pVTZ on H).[22,23] All the high-level ab initio calculations were performed using the CFOUR and MOLPRO program suites.[24,25]

The total atomisation energies at the bottom of the well (TAE$_e$) of the $CH_3CHOO$ and $CH_3CHOO^-$ species are obtained by means of the W3-F12 procedure.[26] W3-F12 theory combines F12 methods[27,28] with extrapolation techniques in order to reproduce the CCSDT(Q) basis set limit energy. The W3-F12 protocol was successfully tested against the W4-11 dataset which also included a selection of molecules exhibiting multi-reference character (i. e. singlet carbenes, radicals, and triplet systems).[26]

Hence, the protocol is deemed suitable to address the non-dynamical correlation effects occurring in the carbonyl oxides.

The CCSD(T)/CBS energy is obtained from the W2-F12 theory and the post-CCSD(T) contributions are obtained from W3.2 theory.[29,30] In brief, the Hartree–Fock component is calculated with the VQZ-F12 basis set (V$n$Z-F12 denotes the cc-pV$n$Z-F12 basis sets of Peterson et al. which were developed for explicitly correlated calculations).[31] Note that the complementary auxiliary basis (CABS) singles correction is included in the SCF energy.[32,33] The valence CCSD-F12 correlation energy is extrapolated from the VTZ-F12 and VQZ-F12 basis sets, using the $E(L) = E_\infty + A/L^\alpha$ two-point extrapolation formula, with $\alpha = 5.94$. In all of the explicitly-correlated coupled cluster calculations the diagonal, fixed-amplitude 3C(FIX) ansatz[32,34,35] and the CCSD-F12b approximation[33,36] are employed. The quasiperturbative triples, (T), corrections are obtained from standard CCSD(T) calculations (i.e., without inclusion of F12 terms) and scaled by the factor $f = 0.987 \cdot E^{MP2-F12}/E^{MP2}$. This approach has been shown to accelerate the basis set convergence.[26,36]

The higher-order connected triples, $T_3 - (T)$, valence correlation contribution is extrapolated from the cc-pVDZ and cc-pVTZ basis sets using the above two-point extrapolation formula with $\alpha = 3$, and the parenthetical connected quadruples contribution (CCSDT(Q)−CCSDT) is calculated with the cc-pVDZ basis set.[29] The CCSD inner-shell contribution is calculated with the core-valence weighted correlation-consistent A′PWCVTZ basis set of Peterson and Dunning,[37] whilst the (T) inner-shell contribution is calculated with the PWCVTZ(no $f$) basis set (where A′PWCVTZ indicates the combination of the cc-pVTZ basis set on hydrogen and the aug-cc-pwCVTZ basis set on carbon, and PWCVTZ(no $f$) indicates the cc-pwCVTZ basis set without the $f$ functions).[26]

The scalar relativistic contribution (in the second-order Douglas–Kroll–Hess approximation[38,39]) is obtained as the difference between non-relativistic CCSD(T)/A′VDZ and relativistic CCSD(T)/A′VDZ-DK calculations (where A′VDZ-DK indicates the combination of the cc-pVDZ-DK basis set on H and aug-cc-pVDZ-DK basis set on C and O).[40] The atomic spin-orbit coupling terms are taken from the experimental fine structure, and the diagonal Born–Oppenheimer corrections (DBOC) are calculated at the HF/cc-pVTZ level of theory. The zero-point vibrational energies (ZPVEs) are derived from the harmonic frequencies (calculated at the CCSD(T)/A′VTZ level of theory for the $CH_3CHOO$ and $CH_3CHOO^-$ species.

The total atomisation energies at 0 K ($TAE_0$) are converted to a heats of formation at 298 K using the Active Thermochemical Tables (ATcT)[41–43] atomic heats of formation at 0 K (H 216.034(1) kJ mol$^{-1}$, C 711.38(6) kJ mol$^{-1}$, and O 246.844(2) kJ mol$^{-1}$), and the CODATA[44] enthalpy functions, $H_{298} - H_0$, for the elemental reference states ($H_2$(g) = 8.468(1) kJ mol$^{-1}$ and C($cr$,graphite) = 1.050(20) kJ mol$^{-1}$), while the enthalpy functions for the $CH_3CHOO$ and $CH_3CHOO^-$ species are obtained within the ridged rotor harmonic oscillator (RRHO) approximation from B3LYP/A′VTZ geometries and harmonic frequencies.[45]

Anion photoelectron spectra were simulated by determining the Franck–Condon Factors (FCFs) linking the anion and neutral $CH_3CHOO$ species vibrational states. FCFs were calculated using the ezSpectrum 3.0 program which is made freely available by Mozhayskiy and Krylov.[46] The program produces FCFs by undertaking Duschinsky rotations of the normal modes between states. Input to the code consists of the output from the ab initio calculations, being geometries, vibrational frequencies, and vibrational normal mode vectors. The predicted stick spectra were convoluted with a Gaussian response function of width 0.002 eV to simulate an experimental spectrum.

## Results and Discussion

### Geometries and vibrational frequencies

In Figure 1, the four ethanal-oxide structures are presented. The neutral, closed-shell equilibrium isomers, both feature $C_S$ symmetry and, as shown in the natural bond orbital (NBO) analysis (Table S2 of the supporting information), they both exhibit a C=O double bond. Hence the neutrals are labelled as *cis*-ethanal-oxide (EO1) and *trans*-ethanal-oxide (EO2). Due to the electrostatic interaction between $H_c$ and $H_d$ atom of the methyl group with the $O_b$ atom, EO1 is by 14.1 kJ mol$^{-1}$ more stable than EO2. A partial charge representation of a natural population analysis is illustrated in Figure S1 of the supporting information.

Contrary to the neutrals, the anionic structures do not show any symmetry, which can be seen when comparing the dihedral angles $\phi_1$ and $\phi_2$ in Figure 1. For this reason and the fact that according to an NBO analysis the anionic structures do not feature a 'classic' C=O double bond, the molecules are labelled as *syn*-ethanal-oxide (AEO1) and *anti*-ethanal-oxide (AEO2), respectively. It is also noticed that the AEO2 isomer is slightly more stable (by 0.4 kJ mol$^{-1}$) than the AEO1 isomer. For both anionic structures, the C−C bond is lengthened on average by 2 pm, the C−O bond by 7 pm and the O−O bond by 10 pm, compared to their neutral counterparts.

Both calculated neutral geometries are in good agreement with the CCSD(T)-F12a/aug-cc-pVTZ structures reported by Nakajima and Endo.[21] The bond lengths reported by the group in the most extreme case deviate by 12 pm (C−C bond), but on average less than 2 pm. It is also noticed that their experimentally determined rotational constants $A$, $B$ and $C$ agree very well with our computed values of $A$ = 17 480 MHz, $B$ = 7139 MHz and $C$ = 5230 MHz. A B3LYP/6-31G(d,p) structure for EO1 reported by Gutbrod and co-workers,[5] also matches the structure we report (where the largest difference is the C−O−O angle, deviating by 1.0°).

Both, the EO1 and EO2 geometries also compare well to the neutral form of the simplest carbonyl oxide, methanal-oxide (MO), where the CCSD(T)/A′VQZ geometry was taken from Ref. [20]), which shows very similar bond lengths [$r$(C−O) = 1.270 Å, $r$(O−O) = 1.343 Å and $\theta$(C−O−O) = 117.9°].

To the best of our knowledge, the is the first published geometry of AEO1 and AEO2. Both however display similar structural features as the anionic form of methanal-oxide (AMO), which exhibits bond lengths of $r$(C−O) = 1.334 Å and $r$(O−O) = 1.450 Å; and bond angles of $\theta$(C−O−O) = 111.7°, $\phi_1$ = 164.8° and $\phi_2$ = −17.9°.[20]

The NBO analysis suggests that the anionic structures harbour the additional electron in the $C_b$ lone-pair orbital, which explains the change from $C_1$ to $C_S$ symmetry when going from the neutral to the anion. Additionally, the C$_a$–C$_b$ bond is weakened significantly as the summed occupancies of these double bonding orbitals drop from ca. 4.0 for the neutrals to ca. 3.0 in the anions. The remaining electron populates a O$_a$ lone pair orbital, which also reveals why the O$_a$–O$_b$ bond is longer for the anion. It should be noted however that the C$_b$–O$_a$ bond orbital and the O$_a$ lone pair orbital are almost degenerate. Compared to the MO and AMO electronic structures, the neutrals and anions are very similar; yet due to the methyl group, the C$_b$–O$_a$ anti-bonding orbitals in the ethanal-oxide species display slightly higher electron density. The detailed NBO analysis and comparison between the carbonyl oxide structures can be found in Table S2 of the supporting information.

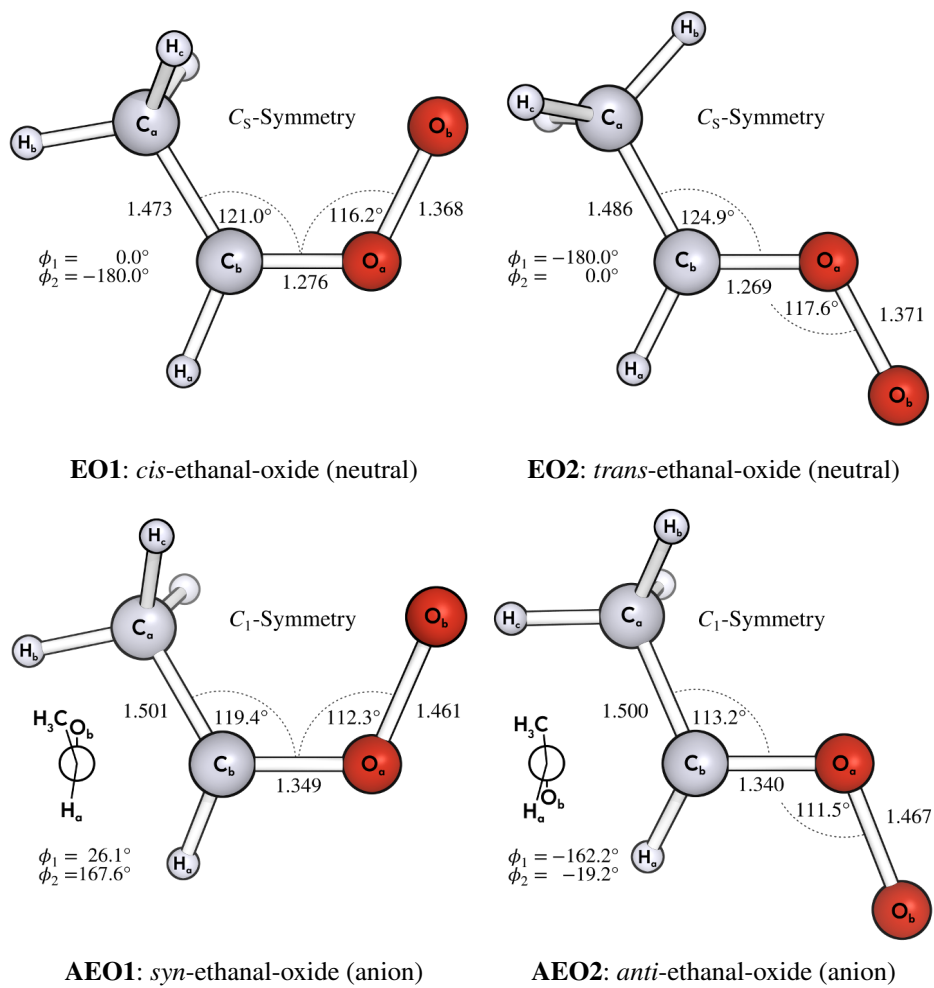

**Figure 1** | Anion and neutral species of CH$_3$CHOO. Bond lengths are given in Ångström. $\phi_1$ refers to the dihedral angle between C$_a$, C$_b$, O$_a$ and O$_b$; $\phi_2$ between H$_a$, C$_b$, O$_a$ and O$_b$. All geometric parameters obtained from CCSD(T)/A′VTZ calculations. The full geometric data are presented in Table S1 of the supplementary information.

The vibrational frequencies are presented in Table 1, where they are sorted in accordance to the standard numbering scheme for the $C_S$ point group. It is possible to approximate the (anharmonic) fundamental frequencies by means of applying a linear scaling factor to the harmonic frequencies.[47] These fundamental frequencies are presented in Table S3 of the supporting information. The idea behind a linear scaling factor is that anharmonic frequencies are in general smaller than their harmonic approximates. In the present case for example the O–O stretching frequency for EO1 is reduced by ca. 33 cm$^{-1}$ from 905 cm$^{-1}$ to 872 cm$^{-1}$ after scaling.

In the neutral structures, EO1 exhibits a weak interaction of the terminal oxygen O$_b$ and the methyl hydrogen

**Table 1** | Computed harmonic vibrational frequencies of the $CH_3CHOO$ anion and neutral species at the CCSD(T)/A'VTZ level of theory. The ordering of the modes of all other species has been changed to reflect that of EO1, for direct comparison. An approximation of the (anharmonic) fundamentals is presented in Table S3 of the supporting information.

| Mode | Neutral | | | Anion | | | Mode description |
|---|---|---|---|---|---|---|---|
| | EO1 | EO2 | Sym. | AEO1 | AEO2 | Sym. | |
| $\nu_1$ | 3201.3 | 3165.7 | $a'$ | 3050.8 | 3104.4 | $a$ | $C_b-H_a$ stretch |
| $\nu_2$ | 3155.4 | 3146.7 | $a'$ | 3090.2 | 3015.3 | $a$ | $C_a-H$ asymmetric stretch (methyl group) |
| $\nu_3$ | 3025.7 | 3034.5 | $a'$ | 2918.0 | 2886.1 | $a$ | $C_a-H$ symmetric stretches (methyl group) |
| $\nu_4$ | 1514.2 | 1524.5 | $a'$ | 1492.5 | 1381.1 | $a$ | asymmetric O−C/C−O stretch |
| $\nu_5$ | 1469.8 | 1461.8 | $a'$ | 1401.6 | 1500.2 | $a$ | $CH_2$ scissor / $H_b-C_aC_b$ bend |
| $\nu_6$ | 1397.0 | 1418.8 | $a'$ | 1361.5 | 1348.4 | $a$ | methyl inversion / C−O stretch |
| $\nu_7$ | 1305.9 | 1312.3 | $a'$ | 1230.7 | 1245.2 | $a$ | in-plane $C_b-H_a$ rocking/ C−O stretch |
| $\nu_8$ | 1113.4 | 1155.2 | $a'$ | 1117.9 | 1118.4 | $a$ | methyl wagging / C−C stretch |
| $\nu_9$ | 973.8 | 942.4 | $a'$ | 913.3 | 937.7 | $a$ | C−C stretch |
| $\nu_{10}$ | 904.9 | 894.8 | $a'$ | 786.8 | 784.8 | $a$ | O−O stretch |
| $\nu_{11}$ | 668.8 | 553.3 | $a'$ | 526.3 | 477.7 | $a$ | methyl wagging / O−O−C bend |
| $\nu_{12}$ | 303.9 | 319.9 | $a'$ | 220.6 | 302.3 | $a$ | C−C−O−O scissor bend/deformation |
| $\nu_{13}$ | 3079.4 | 3097.6 | $a''$ | 3018.2 | 3064.8 | $a$ | $C_a-H$ asymmetric stretch (methyl group) |
| $\nu_{14}$ | 1454.1 | 1479.9 | $a''$ | 1445.9 | 1466.8 | $a$ | $CH_2$ twist / $C_a-H_b$ rocking |
| $\nu_{15}$ | 1034.1 | 1053.5 | $a''$ | 1025.6 | 1029.6 | $a$ | methyl twisting / $C_b-H_a$ rocking |
| $\nu_{16}$ | 723.7 | 843.4 | $a''$ | 648.6 | 654.2 | $a$ | $C_b-H_a$ wagging / methyl rocking |
| $\nu_{17}$ | 442.9 | 247.8 | $a''$ | 305.6 | 120.4 | $a$ | C−C−O−O out-of-plane twisting |
| $\nu_{18}$ | 209.0 | 153.4 | $a''$ | 159.0 | 195.5 | $a$ | methyl twisting (internal rotation) |

All frequencies are given in $cm^{-1}$.

atoms $H_c$ and $H_d$. Similar to a weak hydrogen bond, this interaction stabilises the molecule by 14.1 kJ mol$^{-1}$, compared to EO2. This is also the reason why all the vibrations where $O_b$, $H_c$ and $H_d$ are involved are red-shifted in EO2. The most notable case is the C−C−O−O out-of-plane twisting ($\nu_{17}$) which shows an eigenfrequency of 443 cm$^{-1}$ in EO1 and 248 cm$^{-1}$ in EO2. Another example is the methyl wagging / O−O−C bend ($\nu_{11}$) where the red-shift amounts to 116 cm$^{-1}$. In contrast, the $C_b-H_a$ wagging/methyl rocking vibration ($\nu_{16}$) is blue-shifted by 120 cm$^{-1}$.

The differences in mode frequencies between the anionic and neutral seem quite similar to the differences we encountered in the AMO. Reflecting the increased stability, the O−O stretches ($\nu_{10}$) blue-shift on average by 114 cm$^{-1}$, when comparing the anion to the neutral. It can be seen however, that the blue-shift for the asymmetric O−C / C−O stretch ($\nu_4$) is much larger when going from AEO2 to EO2 than when going from AEO1 to EO1.

## W3-F12 components and Electron affinity

All components of the W3-F12 total atomisation energies for the ethanal-oxide species are given in Table S4 of the supporting information. At the W2-F12 level, the relativistic, all-electron CCSD(T) contributions to TAE$_0$ add up to 2746.5 kJ mol$^{-1}$ (EO1), 2724.0 kJ mol$^{-1}$ (AEO1), 2747.2 kJ mol$^{-1}$ (EO2) and 2709.5 kJ mol$^{-1}$ (AEO2). The generally good performance of the CCSD(T)/CBS level of theory in computational thermochemistry can typically be attributed to the large degree of cancellation between higher-order triples contributions, $T_3 - (T)$, and post-CCSDT contributions. For systems dominated by dynamical correlation, these contributions are of similar magnitude, however, the $T_3 - (T)$ excitations tend to universally decrease the atomisation energies whereas the post-CCSDT excitations tend to universally increase the atomisation energies. An appended Table S5 provides a number of diagnostics for the importance of nondynamical correlation effects, namely the percentage of the total atomisation energy accounted for by the SCF and (T) triples contributions from W2-F12 theory,[29,30] as well as the coupled cluster $T_1$ and $D_1$ diagnostics.[48,49] The $CH_3CHOO$ neutral and anion species considered in the present study exhibit mild-to-moderate nondynamical correlation effects; 64 % to 66 % of the atomisation energy is accounted for at the SCF level, and 2.7 % to 3.1 % by the perturbative triples. The $T_1$ diagnostics of 0.029 to 0.040 and $D_1$ diagnostics of 0.120 to 0.181 also indicate that post-CCSD(T) excitations may have nontrivial contributions. We therefore obtain the CCSDT and CCSDT(Q) contributions from W3-F12 theory. The overall post-CCSD(T) contribution to the atomisation energy amount to 5.3 kJ mol$^{-1}$ and 5.7 kJ mol$^{-1}$ in the neutral structures (EO1 and EO2 respectively), and to 1.6 kJ mol$^{-1}$ and 1.9 kJ mol$^{-1}$ in the anions (AEO1 and AEO2 respectively). We note that the $T_4 - (Q)$, is likely to reduce the magnitude of the connected quadruple excitations, and therefore our CCSDT(Q)/CBS values should be regarded as upper limits of the TAEs.

The heats of formation for the neutral species at 0 K amount to 51.3 kJ mol$^{-1}$ (EO1) and 65.4 kJ mol$^{-1}$ (EO2) At 298 K we report heats of formation of 37.8 kJ mol$^{-1}$ (EO1) and 52.6 kJ mol$^{-1}$ (EO2). For the anionic structures the following heats of formation were obtained: 32.2 kJ mol$^{-1}$ (AEO1) and 31.7 kJ mol$^{-1}$ (AEO2) at 0 K; as well as 20.1 kJ mol$^{-1}$ (AEO1) and 20.0 kJ mol$^{-1}$ (AEO2) at 298 K.

Using the W3-F12 heats of formation for the CH$_3$CHOO neutral and anion species, we were able to calculate the anion electron affinities, the components of which are presented in Table 2. For the EO1 ← AEO1 transition, the electron affinity amounts to 19.1 kJ mol$^{-1}$ at 0 K and 17.7 kJ mol$^{-1}$ at 298 K; for the EO2 ← AEO2 transition, we found an electron affinity of 33.6 kJ mol$^{-1}$ at 0 K and 32.6 kJ mol$^{-1}$ at 298 K. It is noted that the post-CCSD(T) contributions to the electron affinity add up to as much as 3.3 kJ mol$^{-1}$ and 4.1 kJ mol$^{-1}$ for the EO1 ← AEO1 and EO2 ← AEO2 transitions, respectively.

## Predicted anion photoelectron spectra

If the anionic forms of the ethanal-oxide can be synthesised, it should be possible to measure their anionic photoelectron spectra. The main transitions should be the EO1 ← AEO1 transition (T1) and the EO2 ← AEO2 transition (T2). These spectra were simulated employing the ezSpectrum 3.0 code using the geometries, vibrational frequencies, and normal mode vectors of a

Table 2 | Component breakdown of the W3-F12 electron affinities $E_{EA}$ of the ethanal-oxide oxide isomers.

| Property | $E_{EA}$(EO1 ← AEO1) | $E_{EA}$(EO2 ← AEO2) |
|---|---|---|
| SCF | −17.3 | −7.9 |
| CCSD | 41.7 | 46.8 |
| (T) | −8.4 | −7.1 |
| $T_3 - (T)$ | 1.1 | 0.6 |
| (Q) | −4.4 | −4.7 |
| Inner-Shell | −0.6 | −0.5 |
| Rel. | −0.3 | −0.3 |
| DBOC | −0.3 | −0.2 |
| TAE$_e$ | 11.5 | 26.6 |
| ZPVE | −7.6 | −7.0 |
| TAE$_0$ | 19.1 | 33.6 |
| $\Delta H_{f,0}$ | 19.1 | 33.6 |
| $\Delta H_{f,298}$ | 17.7 | 32.6 |

All energies are given in kJ mol$^{-1}$.

CCSD(T)/A′VTZ harmonic calculation. For both simulations, the temperature was set to 10 K, which is appropriate for species entrained in a molecular beam produced via supersonic expansion. Up to 10 quanta were allowed in each excited state vibrational mode (i.e. the modes of the neutral CH$_3$CHOO species). The predicted photoelectron spectra, applying the Duschinsky approach, are presented in Figure 2, where the grey part of the spectrum marks the combination bands, whereas the red part represents the pure progressions; together they form the fully predicted spectrum. To provide a clearer picture of what an experimental spectrum might look like, we the convoluted the stick spectra with a Gaussian response function whose full width at half maximum was set to 0.002 eV and the resulting simulated spectrum is shown in Figure S2 of the supporting information.

During the T1 transition, the geometry only changes slightly and most of the normal modes can be cast onto the neutral system. The determinant of the normal modes rotation matrix, |Det($S$)|, is 0.98 ($S$ is described in the manual of Ref. [46]). For the T2 transition, |Det($S$)| is 0.88, yielding much lower transition intensities (on the order of 10$^2$ times lower).

Most of the the pure T1 transitions are due to $\nu_{17}$, the C−C−O−O out-of-plane twisting mode. Other main contributors to the T1 spectrum are the O−O stretch mode $\nu_{10}$, methyl twisting / C$_b$−H$_a$ rocking mode $\nu_{15}$, C$_b$−H$_a$ wagging / methyl rocking mode $\nu_{16}$ and the methyl twisting mode $\nu_{18}$. The single most important pure progres-

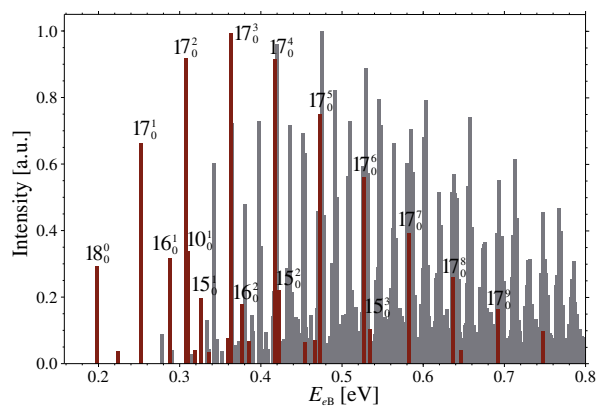

**Figure 2** | Predicted anion photoelectron stick spectra (red marks the pure progressions) for both transitions (Simulated using the Duschinsky approach). Here, $E_{eB}$ is the electron binding energy.

**T1:** EO1 ← AEO1

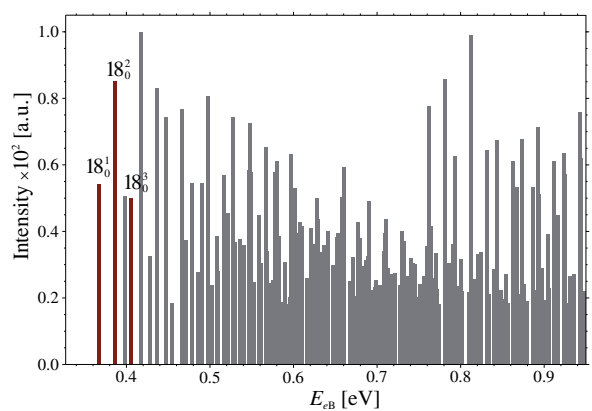

**T2:** EO2 ← AEO2

sion in the T2 spectrum is the methyl twisting mode $\nu_{18}$; this methyl group is rotated during the transition. In general, these observations are consistent with the shifts in vibrational frequencies and changes in geometry during the transition.

## Conclusion

In summary, we show that two different anions, a *syn* and *anti* form, of the ethanal-oxide species exist. They both have a similar geometry and a somewhat similar vibrational structure compared to their corresponding neutral counterparts. By means of the high-level W3-F12 thermochemical protocol we found the electron to be bound by 0.20 eV and 0.35 eV, for the *cis*- and *anti*-ethanal-oxide anion, respectively. While the neutral *cis* and *trans* isomers are separated by 14.1 kJ mol$^{-1}$ (the *cis* form is more stable than the *trans* isomer), the anions are separated by an negligible amount of 0.4 kJ mol$^{-1}$ (here the *anti* form is more stable). The major geometric differences between the anion and neutral are the increased C−O and O−O bond lengths, as well as the change of symmetry, from a $C_S$ point group for the neutrals to $C_1$ in the anions. Anion photoelectron spectra of transitions between these structures were also simulated. Combined with our IR-frequencies, these spectra provide a sound foundation for an analysis of ethanal-oxide structures in future gas-phase experiments.

## Acknowledgements

We gratefully acknowledge an Australian Research Council (ARC) Discovery Early Career Researcher Award (to A.K., project number: DE140 100 311) and the generous allocation of computing time from the National Computational Infrastructure (NCI) National Facility. MK acknowledges the support of an Australian Postgraduate Award (APA). Dr. Alexander Gentleman is acknowledged for his fruitful discussions on the use of the eZ-spectrum software.

## References

[1] R. Criegee. "Mechanism of Ozonolysis". In: *Angewandte Chemie International Edition in English* 14.11 (1975), pp. 745–752. ISSN: 1521-3773. DOI: 10.1002/anie.197507451.

[2] G. T. Drozd, J. Kroll, and N. M. Donahue. "2,3-Dimethyl-2-butene (TME) Ozonolysis: Pressure Dependence of Stabilized Criegee Intermediates and Evidence of Stabilized Vinyl Hydroperoxides". In: *The Journal of Physical Chemistry A* 115.2 (2011), pp. 161–166. DOI: 10.1021/jp108773d.

[3] D. Johnson and G. Marston. "The gas-phase ozonolysis of unsaturated volatile organic compounds in the troposphere". In: *Chemical Society Reviews* 37 (4 2008), pp. 699–716. DOI: 10.1039/B704260B.

[4] R. Gutbrod, R. N. Schindler, E. Kraka, and D. Cremer. "Formation of OH radicals in the gas phase ozonolysis of alkenes: the unexpected role of carbonyl oxides". In: *Chemical Physics Letters* 252.3-4 (Apr. 1996), pp. 221–229. ISSN: 0009-2614. DOI: 10.1016/0009-2614(96)00126-1.

[5] R. Gutbrod, E. Kraka, R. N. Schindler, and D. Cremer. "Kinetic and Theoretical Investigation of the Gas-Phase Ozonolysis of Isoprene: Carbonyl Oxides as an Important Source for OH Radicals in the Atmosphere". In: *Journal of the American Chemical Society* 119.31 (1997), pp. 7330–7342. DOI: 10.1021/ja970050c.

[6] Y.-T. Su, H.-Y. Lin, R. Putikam, H. Matsui, M. C. Lin, and Y.-P. Lee. "Extremely rapid self-reaction of the simplest Criegee intermediate $CH_2OO$ and its implications in atmospheric chemistry". In: *Nature Chemistry* 6.6 (Mar. 2014), pp. 477–483. DOI: 10.1038/nchem.1890.

[7] O. Welz, J. D. Savee, D. L. Osborn, S. S. Vasu, C. J. Percival, D. E. Shallcross, and C. A. Taatjes. "Direct Kinetic Measurements of Criegee Intermediate ($CH_2OO$) Formed by Reaction of $CH_2I$ with $O_2$". In: *Science* 335.6065 (2012), pp. 204–207. DOI: 10.1126/science.1213229.

[8] H. G. Kjærgaard, T. Kurtén, L. B. Nielsen, S. Jørgensen, and P. O. Wennberg. "Criegee Intermediates React with Ozone". In: *The Journal of Physical Chemistry Letters* 4.15 (2013), pp. 2525–2529. DOI: 10.1021/jz401205m.

[9] T. Berndt, T. Jokinen, M. Sipilä, R. L. M. III, H. Herrmann, F. Stratmann, H. Junninen, and M. Kulmala. "$H_2SO_4$ formation from the gas-phase reaction of stabilized Criegee Intermediates with $SO_2$: Influence of water vapour content and temperature". In: *Atmospheric Environment* 89 (2014), pp. 603–612. ISSN: 1352-2310. DOI: 10.1016/j.atmosenv.2014.02.062.

[10] C. A. Taatjes, O. Welz, A. J. Eskola, J. D. Savee, A. M. Scheer, D. E. Shallcross, B. Rotavera, E. P. F. Lee, J. M. Dyke, et al. "Direct Measurements of Conformer-Dependent Reactivity of the Criegee Intermediate $CH_3CHOO$". In: *Science* 340.6129 (2013), pp. 177–180. DOI: 10.1126/science.1234689.

[11] C. A. Taatjes, G. Meloni, T. M. Selby, A. J. Trevitt, D. L. Osborn, C. J. Percival, and D. E. Shallcross. "Direct Observation of the Gas-Phase Criegee Intermediate ($CH_2OO$)". In: *Journal of the American Chemical Society* 130.36 (2008), pp. 11883–11885. DOI: 10.1021/ja804165q.

[12] J. M. Beames, F. Liu, L. Lu, and M. I. Lester. "Ultraviolet Spectrum and Photochemistry of the Simplest Criegee Intermediate CH2OO". In: *Journal of the American Chemical Society* 134.49 (2012), pp. 20045–20048. DOI: 10.1021/ja310603j.

[13] Y.-T. Su, Y.-H. Huang, H. A. Witek, and Y.-P. Lee. "Infrared Absorption Spectrum of the Simplest Criegee Intermediate CH2OO". In: *Science* 340.6129 (Apr. 2013), pp. 174–176. ISSN: 1095-9203. DOI: 10.1126/science.1234369.

[14] L. Vereecken and J. S. Francisco. "Theoretical studies of atmospheric reaction mechanisms in the troposphere". In: *Chemical Society Reviews* 41 (19 2012), pp. 6259–6293. DOI: 10.1039/C2CS35070J.

[15] D.-C. Fang and X.-Y. Fu. "CASSCF and CAS+1 + 2 Studies on the Potential Energy Surface and the Rate Constants for the Reactions between $CH_2$ and $O_2$". In: *The Journal of Physical Chemistry A* 106.12 (2002), pp. 2988–2993. DOI: 10.1021/jp014129m.

[16] M. T. Nguyen, T. L. Nguyen, V. T. Ngan, and H. M. T. Nguyen. "Heats of formation of the Criegee formaldehyde oxide and dioxirane". In: *Chemical Physics Letters* 448.4–6 (2007), pp. 183–188. ISSN: 0009-2614. DOI: 10.1016/j.cplett.2007.10.033.

[17] J. M. L. Martin and G. de Oliveira. "Towards standard methods for benchmark quality ab initio thermochemistry—W1 and W2 theory". In: *The Journal of Chemical Physics* 111.5 (1999), pp. 1843–1856. DOI: 10.1063/1.479454.

[18] A. Arneth, S. P. Harrison, S. Zaehle, K. Tsigaridis, S. Menon, P. J. Bartlein, J. Feichter, A. Korhola, M. Kulmala, et al. "Terrestrial biogeochemical feedbacks in the climate system". In: *Nature Geoscience* 3.8 (Aug. 2010), pp. 525–532. ISSN: 1752-0908. DOI: 10.1038/ngeo905.

[19] N. M. Donahue, G. T. Drozd, S. A. Epstein, A. A. Presto, and J. H. Kroll. "Adventures in Ozoneland: Down the Rabbit-hole". In: *Physical Chemistry Chemical Physics* 13 (23 2011), pp. 10848–10857. DOI: 10.1039/C0CP02564J.

[20] A. Karton, M. Kettner, and D. A. Wild. "Sneaking up on the Criegee intermediate from below: Predicted photoelectron spectrum of the $CH_2OO^-$ anion and W3-F12 electron affinity of $CH_2OO$". In: *Chemical Physics Letters* 585 (Oct. 2013), pp. 15–20. ISSN: 0009-2614. DOI: 10.1016/j.cplett.2013.08.075.

[21] M. Nakajima and Y. Endo. "Communication: Spectroscopic characterization of an alkyl substituted Criegee intermediate syn-$CH_3CHOO$ through pure rotational transitions". In: *The Journal of Chemical Physics* 140.1 (2014), p. 011101. DOI: 10.1063/1.4861494.

[22] T. H. Dunning. "Gaussian basis sets for use in correlated molecular calculations. I. The atoms boron through neon and hydrogen". In: *The Journal of Chemical Physics* 90.2 (1989), pp. 1007–1023. DOI: 10.1063/1.456153.

[23] R. A. Kendall, T. H. Dunning, and R. J. Harrison. "Electron affinities of the first-row atoms revisited. Systematic basis sets and wave functions". In: *The Journal of Chemical Physics* 96.9 (1992), pp. 6796–6806. DOI: 10.1063/1.462569.

[24] CFOUR, a quantum chemical program package written by J.F. Stanton, J. Gauss, M.E. Harding, P.G. Szalay with contributions from A.A. Auer, R.J. Bartlett, U. Benedikt, C. Berger, D.E. Bernholdt, Y.J. Bomble, O. Christiansen, M. Heckert, O. Heun, C. Huber, T.-C. Jagau, D. Jonsson, J. Jusélius, K. Klein, W.J. Lauderdale, D.A. Matthews, T. Metzroth, D.P. O'Neill, D.R. Price, E. Prochnow, K. Ruud, F. Schiffmann, S. Stopkowicz, A. Tajti, J. Vázquez, F. Wang, J.D. Watts and the integral packages MOLECULE (J. Almlöf and P.R. Taylor), PROPS (P.R. Taylor), ABACUS (T. Helgaker, H.J. Aa. Jensen, P. Jørgensen, and J. Olsen), and ECP routines by A. V. Mitin and C. van Wüllen. For the current version, see http://www.cfour.de.

[25] H.-J. Werner, P. J. Knowles, G. Knizia, F. R. Manby, M. Schütz, et al. MOLPRO, *version 2012.1, a package of ab initio programs*. Cardiff, UK, 2012.

[26] A. Karton and J. M. L. Martin. "Explicitly correlated Wn theory: W1-F12 and W2-F12". In: *The Journal of Chemical Physics* 136.12, 124114 (2012), p. 124114. DOI: 10.1063/1.3697678.

[27] K. A. Peterson, D. Feller, and D. A. Dixon. "Chemical accuracy in ab initio thermochemistry and spectroscopy: current strategies and future challenges". In: *Theoretical Chemistry Accounts* 131.1 (2012), pp. 1–20. ISSN: 1432-881X. DOI: 10.1007/s00214-011-1079-5.

[28] S. Ten-no and J. Noga. "Explicitly correlated electronic structure theory from R12/F12 ansätze". In: *Wiley Interdisciplinary Reviews: Computational Molecular Science* 2.1 (2011), pp. 114–125. ISSN: 1759-0884. DOI: 10.1002/wcms.68.

[29] A. Karton, E. Rabinovich, J. M. L. Martin, and B. Ruscic. "W4 theory for computational thermochemistry: In pursuit of confident sub-kJ/mol predictions". In: *The Journal of Chemical Physics* 125.14, 144108 (2006), p. 144108. DOI: 10.1063/1.2348881.

[30] A. Karton, S. Daon, and J. M. Martin. "W4-11: A high-confidence benchmark dataset for computational thermochemistry derived from first-principles W4 data". In: *Chemical Physics Letters* 510.4–6 (2011), pp. 165–178. ISSN: 0009-2614. DOI: 10.1016/j.cplett.2011.05.007.

[31] K. A. Peterson, T. B. Adler, and H.-J. Werner. "Systematically convergent basis sets for explicitly correlated wavefunctions: The atoms H, He, B–Ne, and Al–Ar". In: *The Journal of Chemical Physics* 128.8, 084102 (2008), p. 084102. DOI: 10.1063/1.2831537.

[32] G. Knizia and H.-J. Werner. "Explicitly correlated RMP2 for high-spin open-shell reference states". In: *The Journal of Chemical Physics* 128.15, 154103 (2008), p. 154103. DOI: 10.1063/1.2889388.

[33] T. B. Adler, G. Knizia, and H.-J. Werner. "A simple and efficient CCSD(T)-F12 approximation". In: *The Journal of Chemical Physics* 127.22, 221106 (2007), p. 221106. DOI: 10.1063/1.2817618.

[34] S. Ten-no. "Initiation of explicitly correlated Slater-type geminal theory". In: *Chemical Physics Letters* 398.1–3 (2004), pp. 56–61. ISSN: 0009-2614. DOI: 10.1016/j.cplett.2004.09.041.

[35] H.-J. Werner, G. Knizia, and F. R. Manby. "Explicitly correlated coupled cluster methods with pair-specific geminals". In: *Molecular Physics* 109.3 (2011), pp. 407–417. DOI: 10.1080/00268976.2010.526641.

[36] G. Knizia, T. B. Adler, and H.-J. Werner. "Simplified CCSD(T)-F12 methods: Theory and benchmarks". In: *The Journal of Chemical Physics* 130.5, 054104 (2009), p. 054104. DOI: 10.1063/1.3054300.

[37] K. A. Peterson and T. H. Dunning. "Accurate correlation consistent basis sets for molecular core–valence correlation effects: The second row atoms Al–Ar, and the first row atoms B–Ne revisited". In: *The Journal of Chemical Physics* 117.23 (2002), pp. 10548–10560. DOI: 10.1063/1.1520138.

[38] M. Douglas and N. M. Kroll. "Quantum electrodynamical corrections to the fine structure of helium". In: *Annals of Physics* 82.1 (1974), pp. 89–155. ISSN: 0003-4916. DOI: 10.1016/0003-4916(74)90333-9.

[39] B. A. Hess. "Relativistic electronic-structure calculations employing a two-component no-pair formalism with external-field projection operators". In: *Phys. Rev. A* 33 (6 June 1986), pp. 3742–3748. DOI: 10.1103/PhysRevA.33.3742.

[40] W. A. de Jong, R. J. Harrison, and D. A. Dixon. "Parallel Douglas–Kroll energy and gradients in NWChem: Estimating scalar relativistic effects using Douglas–Kroll contracted basis sets". In: *The Journal of Chemical Physics* 114.1 (2001), pp. 48–53. DOI: 10.1063/1.1329891.

[41] B. Ruscic, R. E. Pinzon, M. L. Morton, G. von Laszevski, S. J. Bittner, S. G. Nijsure, K. A. Amin, M. Minkoff, and A. F. Wagner. "Introduction to Active Thermochemical Tables: Several "Key" Enthalpies of Formation Revisited". In: *The Journal of Physical Chemistry A* 108.45 (2004), pp. 9979–9997. DOI: 10.1021/jp047912y.

[42] W. R. Stevens, B. Ruscic, and T. Baer. "Heats of Formation of $C_6H_5\cdot$, $C_6H_5+$, and $C_6H_5NO$ by Threshold Photoelectron Photoion Coincidence and Active Thermochemical Tables Analysis". In: *The Journal of Physical Chemistry A* 114.50 (2010), pp. 13134–13145. DOI: 10.1021/jp107561s.

[43] B. Ruscic, R. E. Pinzon, M. L. Morton, N. K. Srinivasan, M.-C. Su, J. W. Sutherland, and J. V. Michael. "Active Thermochemical Tables: Accurate Enthalpy of Formation of Hydroperoxyl Radical, $HO_2$". In: *The Journal of Physical Chemistry A* 110.21 (2006), pp. 6592–6601. DOI: 10.1021/jp056311j.

[44] J. D. Cox, D. D. Wagman, and V. A. Medvedev. *CODATA Key Values for Thermodynamics*. see http://www.codata.org. New York, 1989.

[45] P. J. Stephens, F. J. Devlin, C. F. Chabalowski, and M. J. Frisch. "Ab Initio Calculation of Vibrational Absorption and Circular Dichroism Spectra Using Density Functional Force Fields". In: *The Journal of Physical Chemistry* 98.45 (1994), pp. 11623–11627. DOI: 10.1021/j100096a001.

[46] V. Mozhayskiy and A. Krylov. *ezSpectrum*. 2012.

[47] M. K. Kesharwani, B. Brauer, and J. M. L. Martin. "Frequency and Zero-Point Vibrational Energy Scale Factors for Double-Hybrid Density Functionals (and Other Selected Methods): Can Anharmonic Force Fields Be Avoided?" In: *The Journal of Physical Chemistry A* (2014). PMID: 25296165, p. 141021073335006. DOI: 10.1021/jp508422u.

[48] M. L. Leininger, I. M. Nielsen, T. Crawford, and C. L. Janssen. "A new diagnostic for open-shell coupled-cluster theory". In: *Chemical Physics Letters* 328.4–6 (2000), pp. 431–436. ISSN: 0009-2614. DOI: 10.1016/S0009-2614(00)00966-0.

[49] T. J. Lee, J. E. Rice, G. E. Scuseria, and H. F. Schaefer III. "Theoretical investigations of molecules composed only of fluorine, oxygen and nitrogen: determination of the equilibrium structures of FOOF, $(NO)_2$ and FNNF and the transition state structure for FNNF cis-trans isomerization". In: *Theoretica chimica acta* 75.2 (1989), pp. 81–98. ISSN: 0040-5744. DOI: 10.1007/BF00527711.

# Supporting Information

**Table S1 |** CCSD(T) Optimised Geometries at from CCSD(T)/A'VTZ calculation.

| | **EO1**: neutral *cis*-ethanal-oxide | | |
|---|---|---|---|
| | *x* | *y* | *z* |
| C | −0.597 673 | −0.679 246 | 0.000 000 |
| H | −0.958 486 | −1.701 138 | 0.000 000 |
| O | 0.676 240 | −0.604 866 | 0.000 000 |
| O | 1.207 071 | 0.655 454 | 0.000 000 |
| C | −1.427 923 | 0.537 996 | 0.000 000 |
| H | −2.486 964 | 0.288 480 | 0.000 000 |
| H | −1.162 820 | 1.145 816 | 0.871 687 |
| H | −1.162 820 | 1.145 816 | −0.871 687 |
| | **EO2**: neutral *trans*-ethanal-oxide | | |
| | *x* | *y* | *z* |
| C | −0.489 485 | −0.428 951 | 0.000 000 |
| H | −0.216 911 | −1.481 387 | 0.000 000 |
| O | 0.469 175 | 0.402 657 | 0.000 000 |
| O | 1.744 408 | −0.099 797 | 0.000 000 |
| C | −1.879 766 | 0.095 051 | 0.000 000 |
| H | −1.873 391 | 1.184 336 | 0.000 000 |
| H | −2.415 299 | −0.266 933 | 0.881 923 |
| H | −2.415 299 | −0.266 933 | −0.881 923 |
| | **AEO1**: anionic *syn*-ethanal-oxide | | |
| | *x* | *y* | *z* |
| C | −0.634 608 | 0.682 744 | −0.158 401 |
| H | −1.056 198 | 1.638 224 | 0.165 511 |
| O | 0.689 823 | 0.656 521 | 0.095 975 |
| O | 1.253 705 | −0.684 530 | −0.042 937 |
| C | −1.449 240 | −0.560 551 | 0.047 675 |
| H | −2.511 168 | −0.320 638 | −0.089 150 |
| H | −1.159 227 | −1.342 463 | −0.656 091 |
| H | −1.306 578 | −0.985 528 | 1.056 384 |
| | **AEO2**: anionic *anti*-ethanal-oxide | | |
| | *x* | *y* | *z* |
| C | −0.504 603 | 0.462 235 | −0.166 988 |
| H | −0.220 527 | 1.444 238 | 0.209 318 |
| O | 0.452 856 | −0.471 095 | −0.073 743 |
| O | 1.780 206 | 0.133 052 | 0.087 515 |
| C | −1.885 458 | −0.077 265 | 0.091 903 |
| H | −2.636 575 | 0.695 245 | −0.103 910 |
| H | −2.101 140 | −0.929 797 | −0.561 805 |
| H | −2.024 023 | −0.428 465 | 1.131 854 |

Cartesian Coordinates in Å

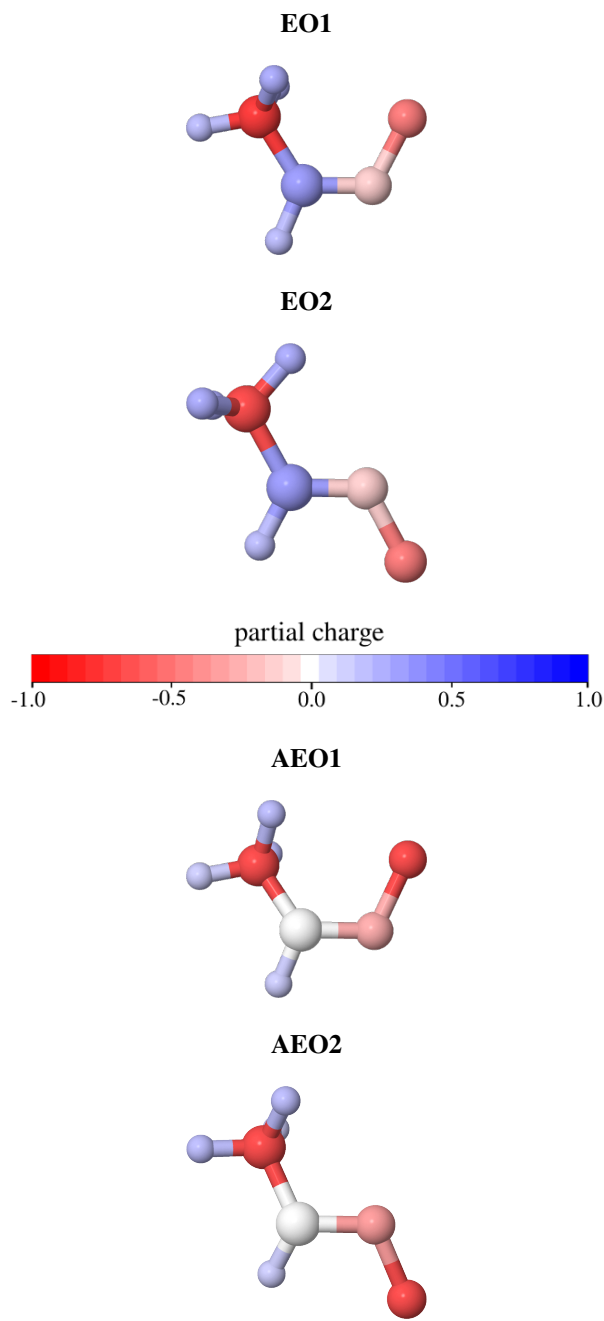

**Figure S1 |** Partial charges sourced from a natural population analysis calculation and projected onto the four ethanal-oxide oxide structures.

**Table S2 |** Orbital occupancies from NBO 6.0 calculations for the carbonyle oxide structures.

| Molecule | EO1 | EO2 | MO* | AEO1 | AEO2 | AMO* |
|---|---|---|---|---|---|---|
| $O_a$ lone pair | 1.9718 | 1.9728 | 1.9780 | 2.9504 | 2.9474 | 2.9706 |
| $O_b$ lone pair | 5.6964 | 5.7164 | 5.8510 | 5.8739 | 5.8978 | 5.9699 |
| $C_a$ lone pair | 0.0000 | 0.0000 | 0.0000 | 0.9277 | 0.9149 | 0.9933 |
| $C_a-O_a$ bonding | 3.9871 | 3.9848 | 3.9959 | 2.9866 | 2.9868 | 2.9960 |
| $C_a-C_b$ bonding | 1.9953 | 1.9855 | – | 1.9947 | 1.9860 | – |
| $O_a-O_b$ bonding | 1.9884 | 1.9852 | 1.9913 | 1.9872 | 1.9822 | 1.9873 |
| $C_a-O_b$ anti-bonding | 0.3453 | 0.3158 | 0.1316 | 0.1214 | 0.0972 | 0.0180 |
| $C_a-C_b$ anti-bonding | 0.0190 | 0.0126 | – | 0.0211 | 0.0148 | – |
| $O_a-O_b$ anti-bonding | 0.0188 | 0.0159 | 0.0085 | 0.0183 | 0.0152 | 0.0088 |

* from Ref. [20]

**Table S3 |** Computed scaled harmonic vibrational frequencies of the $CH_3CHOO$ anion and neutral species at the CCSD(T)/A'VTZ level of theory The scaling factor of 0.9635 taken Kesharwani et al. is the scaling factor for CCSD(T)/cc-pV(T+d)Z calculations (the +d only refers to 2nd row atoms which are not present here). The basis set used in this study is more or less the same; yet we included augmented functions for C and O. This is expected to have a small effect on the scaling factor. The ordering of the modes of all other species has been changed to reflect that of EO1, for direct comparison.

| Mode | Neutral EO1 | Neutral EO2 | Sym. | Anion AEO1 | Anion AEO2 | Sym. | Mode description |
|---|---|---|---|---|---|---|---|
| $\nu_1$ | 3084.5 | 3050.2 | $a'$ | 2939.4 | 2991.1 | $a$ | $C_b-H_a$ stretch |
| $\nu_2$ | 3040.2 | 3031.8 | $a'$ | 2977.4 | 2905.2 | $a$ | $C_a-H$ asymmetric stretch (methyl group) |
| $\nu_3$ | 2915.3 | 2923.7 | $a'$ | 2811.5 | 2780.8 | $a$ | $C_a-H$ symmetric stretches (methyl group) |
| $\nu_4$ | 1458.9 | 1468.9 | $a'$ | 1438.0 | 1330.7 | $a$ | asymmetric O–C/C–O stretch |
| $\nu_5$ | 1416.2 | 1408.4 | $a'$ | 1350.4 | 1445.4 | $a$ | $CH_2$ scissor / $H_b-C_aC_b$ bend |
| $\nu_6$ | 1346.0 | 1367.0 | $a'$ | 1311.8 | 1299.2 | $a$ | methyl inversion / C–O stretch |
| $\nu_7$ | 1258.2 | 1264.4 | $a'$ | 1185.8 | 1199.8 | $a$ | in-plane $C_b-H_a$ rocking/ C–O stretch |
| $\nu_8$ | 1072.8 | 1113.0 | $a'$ | 1077.1 | 1077.6 | $a$ | methyl wagging / C–C stretch |
| $\nu_9$ | 938.3 | 908.0 | $a'$ | 880.0 | 903.5 | $a$ | C–C stretch |
| $\nu_{10}$ | 871.9 | 862.1 | $a'$ | 758.1 | 756.2 | $a$ | O–O stretch |
| $\nu_{11}$ | 644.4 | 533.1 | $a'$ | 507.1 | 460.3 | $a$ | methyl wagging / O–O–C bend |
| $\nu_{12}$ | 292.8 | 308.2 | $a'$ | 212.5 | 291.3 | $a$ | C–C–O–O scissor bend/deformation |
| $\nu_{13}$ | 2967.0 | 2984.5 | $a''$ | 2908.0 | 2952.9 | $a$ | $C_a-H$ asymmetric stretch (methyl group) |
| $\nu_{14}$ | 1401.0 | 1425.9 | $a''$ | 1393.1 | 1413.3 | $a$ | $CH_2$ twist / $C_a-H_b$ rocking |
| $\nu_{15}$ | 996.4 | 1015.0 | $a''$ | 988.2 | 992.0 | $a$ | methyl twisting / $C_b-H_a$ rocking |
| $\nu_{16}$ | 697.3 | 812.6 | $a''$ | 624.9 | 630.3 | $a$ | $C_b-H_a$ wagging / methyl rocking |
| $\nu_{17}$ | 426.7 | 238.8 | $a''$ | 294.4 | 116.0 | $a$ | C–C–O–O out-of-plane twisting |
| $\nu_{18}$ | 201.4 | 147.8 | $a''$ | 153.2 | 188.4 | $a$ | methyl twisting (internal rotation) |

All frequencies are given in $cm^{-1}$.

**Table S4** | Component breakdown of the W3-F12 total atomisation energies and heats of formation of the $CH_3CHOO$ neutral and anion species (in kJ/mol) as well as the percentaged (T) contribution as an indicator for non-dynamic correlation.

| Molecule | AEO1 | EO1 | AEO2 | EO2 |
|---|---|---|---|---|
| SCF | 1860.0 | 1877.3 | 1866.8 | 1874.8 |
| CCSD | 949.6 | 907.9 | 944.1 | 897.2 |
| (T) | 79.8 | 88.3 | 78.8 | 85.9 |
| $T_3 - (T)$ | −3.6 | −4.6 | −3.6 | −4.2 |
| (Q) | 5.5 | 9.9 | 5.2 | 9.9 |
| Inner-Shell | 10.0 | 10.6 | 10.1 | 10.6 |
| Scalar Relativistic | −3.0 | −2.6 | −3.0 | −2.7 |
| Spin-Orbit | −2.6 | −2.6 | −2.6 | −2.6 |
| DBOC | 0.4 | 0.6 | 0.4 | 0.6 |
| $TAE_e$ | 2896.2 | 2884.7 | 2896.2 | 2869.6 |
| ZPVE | 147.8 | 155.4 | 147.3 | 154.4 |
| $TAE_0$ | 2748.4 | 2729.3 | 2748.8 | 2715.2 |
| $\Delta H_{f,0}$ | 32.2 | 51.3 | 31.7 | 65.4 |
| $\Delta H_{f,298}$ | 20.1 | 37.8 | 20.0 | 52.6 |

**Table S5** | Diagnostics for importance of nondynamical correlation.

| Molecule | AEO1 | EO1 | AEO2 | EO2 |
|---|---|---|---|---|
| %TAE[(T)][a] | 2.76 | 3.07 | 2.73 | 3.01 |
| %TAE[SCF][a] | 64.37 | 65.33 | 64.6 | 65.6 |
| $T_1$[b] | 0.031 | 0.04 | 0.029 | 0.039 |
| $D_1$[b] | 0.132 | 0.181 | 0.12 | 0.18 |

a From W2-F12 theory.
b From CCSD-F12/VQZ-F12 calculations.

**Figure S2 |** Predicted Gaussian convoluted anion photoelectron spectra for both transitions (full width at half maximum is 0.002 eV).

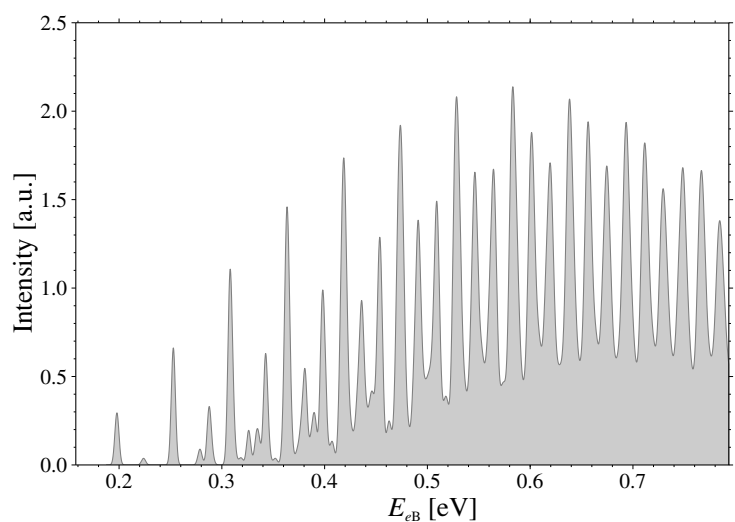

**T1:** EO1 ← AEO1

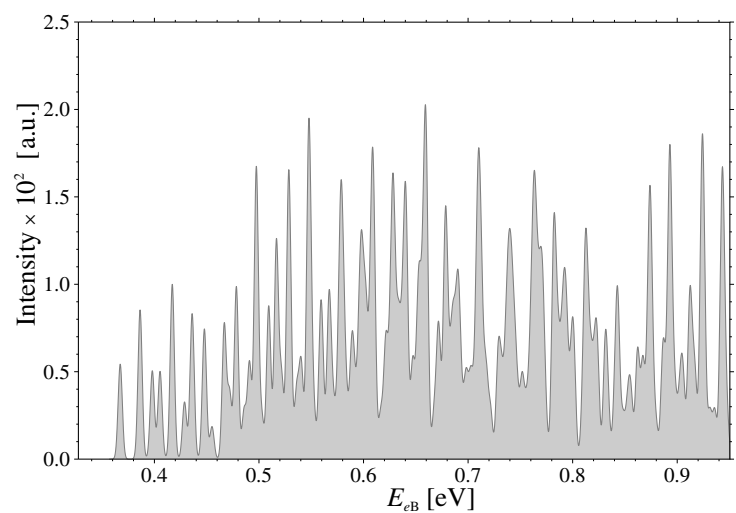

**T2:** EO2 ← AEO2